\documentclass[sn-mathphys,Numbered]{sn-jnl}

\usepackage{graphicx}%
\usepackage{multirow}%
\usepackage{amsmath,amssymb,amsfonts}%
\usepackage{amsthm}%
\usepackage{mathrsfs}%
\usepackage[title]{appendix}%
\usepackage{xcolor}%
\usepackage{textcomp}%
\usepackage{manyfoot}%
\usepackage{booktabs}%
\usepackage{algorithm2e}%
\usepackage{algorithmicx}%
\usepackage{algpseudocode}%
\usepackage{listings}%
\usepackage{subcaption}
\usepackage{multirow}
\usepackage{booktabs}
\usepackage{array}
\usepackage{adjustbox}
\usepackage{rotating}

\usepackage{tcolorbox}

\theoremstyle{thmstyleone}%
\theoremstyle{thmstyletwo}%

\theoremstyle{thmstylethree}%
\begin{document}

\title[Article Title]{Proximal Policy Optimization-Based Reinforcement Learning 
Approach for DC-DC Boost Converter Control: A Comparative Evaluation Against Traditional Control Techniques \footnotemark[2]\footnotetext[2]{This paper has been accepted and published by Elsevier Ltd. (Heliyon)}}

\author[1,2]{\fnm{Utsab} \sur{Saha}}

\author[3]{\fnm{Atik} \sur{Jawad}}

\author[1]{\fnm{Shakib} \sur{Shahria}}

\author*[1]{\fnm{A.B.M Harun-Ur} \sur{Rashid}}\email{abmhrashid@eee.buet.ac.bd}

\affil[1]{\orgdiv{Department of Electrical and Electronic Engineering}, \orgname{Bangldesh University of Engineering and Technology}, \orgaddress{\city{Dhaka}, \postcode{1205}, \country{Bangladesh}}}

\affil[2]{\orgdiv{School of Data and Sciences}, \orgname{BRAC University}, \orgaddress{\city{Dhaka}, \postcode{1212}, \country{Bangladesh}}}

\affil[3]{\orgdiv{Department of Electrical and Electronic Engineering}, \orgname{University of Liberal Arts Bangladesh}, \orgaddress{\city{Dhaka}, \postcode{1207}, \country{Bangladesh}}}


\abstract{This article proposes a proximal policy optimization (PPO)-based reinforcement learning (RL) approach for DC-DC boost converter control that is compared with traditional control methods. 
The performance of the PPO algorithm is evaluated using MATLAB Simulink co-simulation, and 
the results demonstrate that the most efficient approach for achieving short settling time and 
stability is to combine the PPO algorithm with a reinforcement learning-based control method. 
The simulation results show that the control method based on RL with the PPO algorithm provides step response characteristics that outperform traditional control approaches, thereby 
enhancing DC-DC boost converter control. This research also highlights the inherent capability of the reinforcement learning method to enhance the performance of boost converter control.}




\maketitle

\section{Introduction}\label{sec1}
In recent times, DC power systems have been favored over AC power systems in many 
applications due to their superior reliability and quality. DC loads are commonly used in modern electronics, as DC is a preferred option for microgrid (MG) designs \cite{marx2011large}, \cite{singh2016mitigation}. Moreover, there has been 
a significant focus on the production of electricity from sustainable energy sources. These systems require sophisticated control techniques in order to fully utilize their potential. This can involve 
adjusting the load to match the voltage of the source, known as maximum power-point tracking, to 
extract the maximum power. Alternatively, it can involve maintaining a steady power supply for a passive load. As a consequence, the investigation of a boost converter, which is a DC-to-DC converter that controls the voltage between a power supply and a load, has gained popularity. 

The performance of power electronic systems, particularly boost converters, can be greatly 
improved by using proper control methods. The converter output voltage and current are regulated as part of these control schemes to provide the necessary power at the desired condition. Boost converters are widely utilized in a variety of industries, including telecommunications \cite{khursheed2021tuning}, electric vehicles \cite{janabi2019switched, kumar2021performance}, and renewable energy systems \cite{hasanpour2020new, alghaythi2020high} where the main function is to step up the voltage level to fulfill the needed power demand at the specified condition. However, a proper control circuit is essential for guaranteeing dependable performance, increasing efficiency, and extending the system lifespan of the converter. Boost converters are controlled using several control techniques such 
as proportional integral derivative (PID) \cite{ounnas2019design, arulselvi2004design}, artificial neural network (ANN) \cite{dhivya2013neural, koduru2022real}, model 
predictive control (MPC) \cite{kim2014stabilizing}, fuzzy logic \cite{ismail2010fuzzy, bendaoud2017implementation}, adaptive neuro-fuzzy inference system (ANFIS) 
control, sliding mode control \cite{guldemir2005sliding}, etc.

Traditional control techniques, such as PI control, have been widely used due to their simplicity and robustness. PI control works by comparing the output voltage or current with the reference value and adjusting the duty cycle of the converter switch accordingly. However, PI control is a linear 
control technique and may not always be optimal for improving the performance of nonlinear systems 
\cite{aguilera2018basic}. Another popular control algorithm is the adaptive neuro-fuzzy inference system (ANFIS), which is described in \cite{denai2004anfis, saha2023intelligent}. ANFIS is a type of fuzzy logic control that uses a neural network to tune the fuzzy logic system parameters. ANFIS combines the benefits of fuzzy logic and neural networks to 
provide a more accurate and efficient control approach. Artificial neural networks (ANN) is another machine learning technique used in power electronics control \cite{bose2001artificial, zhao2020overview}. ANN is a type of supervised learning that uses a backpropagation algorithm to optimize the control policy. ANN has shown promising results in improving the control of various power electronic systems, including boost converters, but to get an accurate model, it requires a proper dataset, more training time, and a slower speed of learning. Fuzzy logic control is another control technique that uses linguistic rules to control 
nonlinear systems. It works by defining the input and output variables in linguistic terms and using a set of fuzzy rules to map the inputs to the outputs. However, there are a few issues with fuzzy logic 
controllers. Fuzzy control methods rely on human ability and understanding. Defining specific fuzzy sets or membership functions takes time and effort.

Popular power converter control strategies employed for buck-boost converters with constant 
power loads (CPLs) encompass state-feedback controllers \cite{wu2006cascade, mohamed2018control}, proportional-integral-derivative (PID) control \cite{sumita2019pid, kobaku2020quantitative}, model predictive control (MPC) \cite{xu2019offset, boutchich2020constrained}, adaptive feedback controller \cite{zhang2022adaptive} 
and sliding mode control (SMC)~\cite{fan2016gpi, louassaa2023robust, singh2015robust, shen2023cascade, wu2021sliding, liu2022fixed}. Recent research endeavors focusing on voltage regulation and stabilization employing intelligent controllers have explored various techniques. These include the utilization of a quadratic D-stable fuzzy controller \cite{mardani2018design}, fuzzy-PID 
control \cite{bastos2014intelligent}, as well as machine learning and reinforcement learning approaches such as deep 
deterministic policy gradient (DDPG) \cite{9097405}, and deep reinforcement learning (DRL) methods like Markov decision process (MDP) and deep Q network (DQN) algorithms \cite{cui2020intelligent},~\cite{WU2023110999}. Additionally, a modified fast terminal sliding mode control (FTSMC) with a fixed switching frequency has been proposed to regulate the output voltage of DC-DC buck converters \cite{balta2022modified}.

As boost converters are becoming an essential part of DC micro-grids as well as sustainable energy generation (i.e. solar PV systems), the problem associated with these applications should be considered very carefully. In DC microgrids (MGs), there are issues of voltage instability, which is caused by a constant power load (CPL), which has been extensively investigated and documented in existing literature \cite{singh2017mitigation}, \cite{andalibi2021time}. Over time, control techniques addressing this challenge have transitioned from model-based approaches to model-free methods. These techniques involve various controller systems, including traditional state feedback controllers, and more recently, the utilization of machine learning techniques. Based on the reviewed literature, the essential areas of research that need more investigation can be determined as follows:
\begin{itemize}
    \item The existence of CPLs in DC-MGs introduces a significant concern due to their nonlinear 
character.
    \item Model-based strategies may exhibit limitations in effectively managing uncertainties encountered in practical applications.
    \item In the case of neural network control, proper dataset and training time become a big challenge.
    \item In DC-to-DC boost converters, the traditional control method exhibited increased instability and inferior step response characteristics in the presence of changes in dynamics 
in the system.
\end{itemize}

\noindent\textbf{Our Contribution. }A potential solution to address these issues is the implementation of an intelligent controller capable of interacting with the system and learning to effectively handle both normal and abruptly changing 
system dynamics. By adapting and acquiring knowledge through interaction with the system, such intelligent controllers can improve the control performance in various operating conditions. Reinforcement learning can be an effective approach for tackling these difficulties associated with enhancing converter control. Reinforcement learning is a machine learning method that has drawn 
considerable interest in the domain of power electronics control. It has been extensively explored in recent studies \cite{abel2022theory, cao2020reinforcement, hajihosseini2020dc}. RL operates by learning an optimal control policy through iterative interaction with the environment. This approach has shown promising outcomes in improving the control of diverse nonlinear systems, including several power converters. RL is a model-free framework that is capable of solving optimum control problems. The controller in the general feedback control structure receives information about the condition of the system and responds correspondingly. Similarly, the decision rule used in reinforcement learning (RL) is a control law known as the policy \cite{bucsoniu2018reinforcement}. This policy governs the actions taken by the system based on state feedback. The state of the system is changed by the applied actuation, and when it does so, the transition to the updated state is assessed using a reward function. Increasing the cumulative reward from each initial state is the main objective of optimal control. The ultimate objective is to optimize the system’s long-term performance because decision-making in this process is sequential. There are many reliable model-free RL methods. In this paper, a specific algorithm called proximal policy optimization (PPO) is implemented, which directly optimizes the policy parameters using observed data \cite{bucsoniu2018reinforcement, prag2021data}. The purpose of this study is to assist in the research of improving the utilization of renewable energy sources for generating electricity. The goal is to conserve fossil fuel energy resources. This involves applying concepts such as maximum point tracking to boost converters with a dynamic load, or to schemes that maintain a constant output voltage with a fixed resistance load. This work aims to design an adaptive control methodology to address the voltage instability and achieve the target voltage in DC-to-DC boost converters. The goal is to minimize the settling time. The paper’s key contribution can be summarized as follows-
\begin{itemize}
    \item This work presents a method for controlling a boost converter using proximal policy optimization (PPO), an advanced reinforcement learning (RL) methodology known for its effective use in several control applications.
    \item The performance of the PPO-based control approach is compared with traditional control techniques, including optimized proportional-integral (PI) control and artificial neural network (ANN) control.
    \item The effectiveness of the proposed control approach in improving the performance of the boost converter control system is evaluated through simulations using the MATLAB Simulink solver. 
    \item To evaluate the reliability of the control performance, simulations are performed under both fixed and varying input conditions.
\end{itemize}
Through extensive experimentation, it is observed that, in both scenarios, the step response 
characteristics remained consistently similar and almost better than the existing control methods. The subsequent sections of the paper are arranged in the following manner. Section 2 provides an explanation of the fundamental operational mechanism of the boost converter. The proposed methodology is outlined in Section 3, with a detailed algorithm provided to illustrate each step. Section 4 provides a comprehensive account of the experiments and simulation situations. Section 5 
presents the validation of our method and a comparative study with existing studies. Section 6 presents a concise summary of our findings and insights.

\section{Boost Converter Model}\label{sec:boost_model}
\begin{figure}[t]
    \centering
    \includegraphics[width=0.9\textwidth]{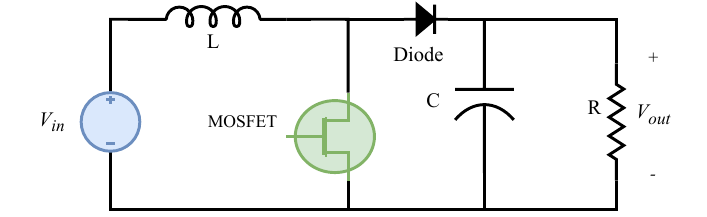}
    \caption{Circuit diagram of DC-DC boost converter}
    \label{fig:1}
\end{figure}

The boost converter operates on the principle of complementary switching. It encompasses two 
complementary modes: the closed mode and the open mode \cite{7086663}. In the closed mode, energy is stored 
in the inductor while being released from the capacitor. Conversely, in the open mode, energy is released from the inductor while being stored in the capacitor. The operation of a boost converter relies on two fundamental principles, namely the energy balance principle and the charge balance principle, which ensure its ideal functioning \cite{7086663}. The circuit diagram of the boost converter is shown in Fig. \ref{fig:1}.

Based on the charge balance and energy balance principles, the basic relationship between the input 
voltage and output voltage of the boost converter is given by Eq. (\ref{first}).
\begin{equation}\label{first}
\mathrm{V}_{out}^{}  =\frac{\mathrm{V}_{in}^{}}{1-d}
\end{equation}
Where ${V}_{in}$ is the input voltage, ${V}_{out}$ is the output voltage and $d$ is the duty cycle.
The averaging method is employed for the analysis and design of controllers in power electronics 
circuits. To utilize the averaging method for modeling purposes, the state space equations for the 
closed and open modal operations are represented by Eq. (\ref{eq:two}) and Eq. (\ref{eq:three}).
\begin{equation}
\label{eq:two}
    \dot{x} = \mathrm{A}_{1}^{}x+\mathrm{B}_{1}^{}u
\end{equation}
Here, $\dot{x}$ is the state variable (inductor current and capacitor voltage) and $u$ is the input voltage, ${V}_{in}$.
\begin{equation}
\label{eq:three}
\dot{x} = \mathrm{A}_{2}^{}x+\mathrm{B}_{2}^{}u
\end{equation}
Then the average state space model is represented by Eq. (\ref{eq:four}).
\begin{equation}
\label{eq:four}
    \dot{x} = \mathrm{A}_{}^{}x+\mathrm{{B}}_{}^{}u  
\end{equation}

 where,
 $$ {A} =  \mathrm{A}_{1}^{}d+\mathrm{A}_{2}^{}(1-d)$$
 $$ {B} =  \mathrm{B}_{1}^{}d+\mathrm{B}_{2}^{}(1-d)$$
 $$\mathrm{A}_{1} = \begin{bmatrix}
0 & 0 \\
0  & -\frac{1}{RC}
\end{bmatrix}$$
$$\mathrm{B}_{1}= \mathrm{B}_{2} = \begin{bmatrix}
\frac{1}{L}\\
0
\end{bmatrix}$$
$$\mathrm{A}_{2} = \begin{bmatrix}
 0& -\frac{1}{L} \\
 \frac{1}{C}&-\frac{1}{RC} 
\end{bmatrix}$$

\section{Proposed Methodology
}\label{sec3}
This section depicts the proposed control method for the DC-DC boost converter. This paper introduces an improved method for controlling dc-to-dc boost converters, which combines 
reinforcement learning (RL) with proximal policy optimization (PPO). The proposed boost converter 
control technique is based on a feedback mechanism. It utilizes the observed voltage reading across 
the load or resistance as the feedback signal. The control action is dictated by the state of the 
MOSFET switch in the DC-to-DC boost converter. The speed and efficiency at which the boost converter reaches the specified output voltage are determined by the sequence of control signals transmitted to the MOSFET switch. The subsequent subsections present a comprehensive depiction of 
the proposed method in detail.

\subsection{Control Method Based on Reinforcement Learning}
Reinforcement learning (RL) is a machine learning technique where an agent learns to make optimal decisions by continuously interacting with its environment and maximizing a reward signal. This approach takes inspiration from the learning mechanisms observed in humans and animals, involving a trial-and-error process to acquire and improve decision-making skills. RL is applied in various domains where making effective decisions under challenging conditions is crucial. In practical scenarios, an agent encounters situations where it needs to make decisions, and it relies on feedback from the environment in the form of rewards or penalties to adjust and refine those decisions. To optimize the long-term cumulative reward, the agent must acquire a policy, a set of rules or strategies, that guides its decision-making process. Importantly, the agent does not receive explicit instructions or guidance on how to achieve this objective; instead, it learns by actively engaging with the 
environment, gaining knowledge through repeated attempts, and learning from both successful and unsuccessful outcomes.

Reinforcement learning (RL) is a framework devoid of explicit models, capable of solving optimal control problems. The controller inside the general feedback control structure gets feedback in the form of state signals from the plant and acts accordingly. In RL, the decision rule is denoted as a 
policy, which operates based on state feedback control principles \cite{bucsoniu2018reinforcement}. The system's state is modified through actuation, and the resulting transition to the new state is assessed using a reward function. The objective of optimum control is to maximize the overall reward obtained from each initial state. The purpose of this process, which involves sequential decision-making, is to optimize the system’s long-term performance. Various robust model-free RL algorithms exist, and this paper focuses on a policy gradient method called proximal policy optimization (PPO) \cite{bucsoniu2018reinforcement}. PPO directly optimizes policy parameters using observed data.
\begin{figure}[t]
    \centering
    \includegraphics[width=0.9\textwidth]{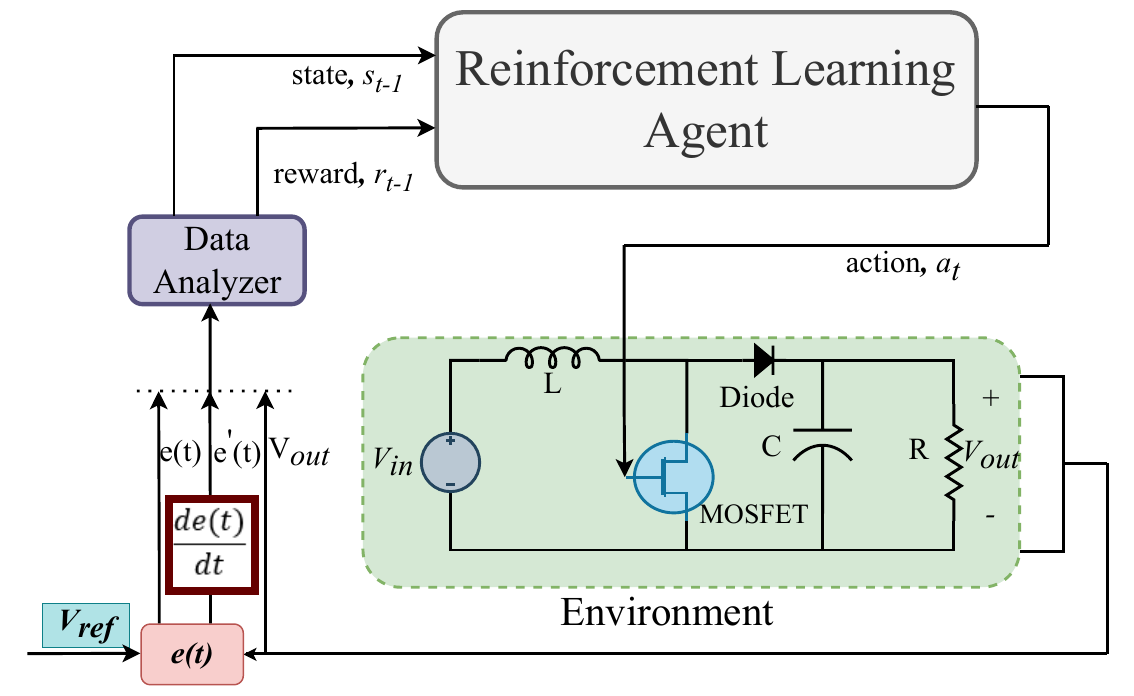}
    \caption{Diagram of proposed method}
    \label{fig:2}
\end{figure}

In this study, the environment corresponds to the DC-DC boost converter system. The action, taken by the agent, involves generating the gate pulse that regulates the converter’s operation as shown in Fig. \ref{fig:2}. The data analyzer block in Fig. \ref{fig:2} continuously monitors important parameters such as the output voltage, $V_{out}$, error (deviation from the desired value), $e\left( t \right)$, and the rate of change of error, $e^{'}\left( t \right)$. These monitored signals are processed as state, $s_{t-1}$, and reward, $r_{t-1}$, and provided to the RL agent, which is shown in Fig. \ref{fig:2}. Using the analyzed state, $s_{t-1}$, and reward information, $r_{t-1}$ the RL agent makes informed decisions regarding the appropriate action (gate pulse) to be applied. The agent leverages its learned policy to select an action, that is expected to optimize the cumulative reward over time. Through a trial-and-error process, the agent explores different gate pulse actions and learns from the feedback received through the reward signal. By repeatedly adjusting its actions based on the observed outcomes, the agent adapts its decision-making strategy, gradually improving its control performance in regulating the boost converter. 

\begin{figure}[t]
    \centering
  \includegraphics[width=0.4\textwidth]{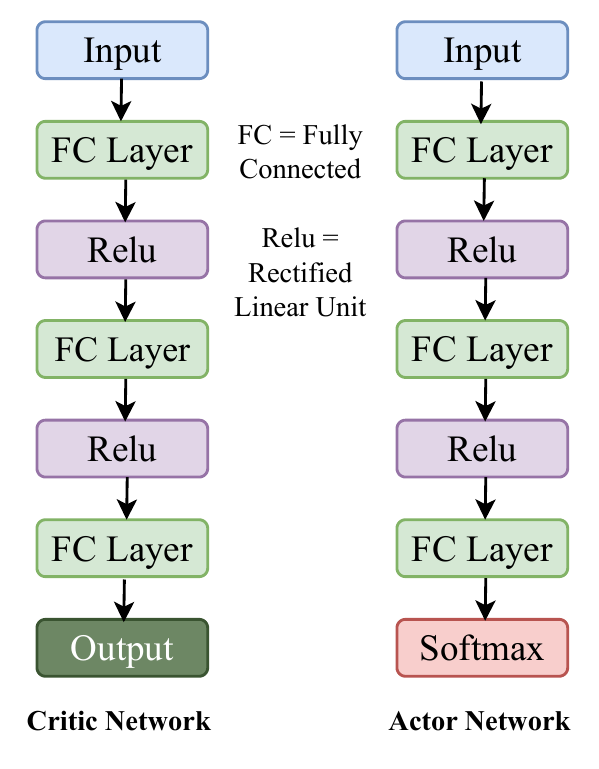}
    \caption{Network architecture of actor-critic network}
    \label{fig:3}
\end{figure}
\subsection{Proximal Policy Optimization Algorithm}
The proximal policy optimization (PPO) algorithm is a model-free, online, or on-policy reinforcement learning (RL) method. It updates the decision-making policy using small batches of experiences obtained from interacting with the environment. PPO keeps a balance between important factors like ease of implementation, tuning, sample complexity, and efficiency while minimizing the deviation from the previous policy. It learns from online data and ensures low variance in training \cite{prag2021data}. In this work, the PPO algorithm follows an iterative process where it alternates between two key steps: sampling data by interacting with the environment which is the boost converter, and optimizing a clipped surrogate objective function using stochastic gradient ascent \cite{schulman2017proximal}. The algorithm ensures stability during agent training by utilizing the clipped surrogate objective function and constraining the magnitude of policy changes at each iteration \cite{liu2019neural}. This approach helps to prevent drastic policy updates and contributes to smoother and more reliable training of the agent. PPO is commonly implemented using the actor-critic model, which consists of two deep neural networks - one for action selection (actor) and the other for reward evaluation (critic). In reinforcement learning, the actor network is responsible for decision-making by selecting actions based on the current state of the environment which is a boost converter in this case. Its goal is to optimize a policy that maps states to actions effectively. On the other hand, the critic network evaluates the actions chosen by the actor network and provides feedback on the value of state-action pairs. This feedback helps the actor network to refine and improve its policy. The detailed architecture of the actor-critic network is shown in Fig. \ref{fig:3}. In this work, to construct the state vector ($s_t$) at each sample time step, the PPO agent calculates the output voltage ($V_{out}$), the error value e(t), and the rate of change of the error $e^{'}\left( t \right)$ as shown in Fig. \ref{fig:2}.

The objective of the actor-critic network is to optimize the surrogate objective function, L$(\theta)$ with the aim of maximizing performance. The expression of L$(\theta)$ is described in Eq. (\ref{eq:five}).
\begin{equation}
\label{eq:five}
L(\theta) = \mathrm{E}_{t}^{'}[min(\mathrm{r}_{t}^{}(\theta))\mathrm{A}_{t}^{'}, clip(\mathrm{r}_{t}^{}(\theta),1-\in ,1+\in)\mathrm{A}_{t}^{'}]    
\end{equation}
This function represents the expectation of the advantage function, where the advantage function is dependent on the estimated advantage $\mathrm{A}_{t}^{}$, the policy parameters $\theta$, and the probability ratio $\mathrm{r}_{t}^{}(\theta)$.
\begin{equation}
\label{eq:six}
    r_t(\theta )=\frac{\pi_{ \theta}\left ( a_t \mid s_t\right )}{\pi_{\theta old }\left ( a_t \mid s_t\right ) }
\end{equation}
The probability ratio, $\mathrm{r}_{t}^{}(\theta)$ in the context of the actor-critic network, refers to the comparison between the likelihood of taking a specific action when in state st at time $t$, based on the current policy parameters $\pi_{ \theta}$, and the likelihood of taking the same action at in the same state $s_t$ at time $t$, utilizing the past or previous policy parameters $\theta_\text{old}$ from the previous epoch, which is described in Eq. \ref{eq:six}. During training, the PPO-based RL agent learns by sampling actions according to its updated policy. This policy starts with random actions to explore the state-action space and gradually becomes more focused on actions that result in higher rewards. The PPO agent determines the probabilities for each action in its set of possible actions. It randomly chooses an action using these probabilities. The agent updates its actor and critic properties using mini-batches of data over multiple training sessions while interacting with the environment. The objective of the PPO agent is to train the coefficients of the actor-critic neural networks to minimize the difference between the desired output $V_{ref}$ and the actual 
value $V_{out}$.

The general algorithmic structure of the PPO algorithm \cite{hajihosseini2020dc},~\cite{prag2021data} is described in Algorithm \ref{alg:ppo}. The PPO algorithm follows a specific set of steps. Initially, the parameters of the actor-critic network are initialized. Next, a sequence of experiences is generated, consisting of state-action pairs and their corresponding reward values. For each time instance, the action-value function, $Q^\pi$, and advantage function, $A^\pi$ are calculated. The action-value function represents the expected return when starting 
from a particular state and taking a specific action according to the policy. It is computed as the sum of the expected rewards associated with the state-action pair. On the other hand, the value function estimates the expected return of being in a particular state and reflects its desirability. It is computed as 
the sum of the expected rewards given the state as shown in Algorithm \ref{alg:ppo}. The advantage function, $A^\pi$ captures the difference between the action-value function and the value function. It provides insights into the advantages or disadvantages of choosing a specific action in a given state. During the training 
process, the PPO algorithm learns from a set of mini-batch experiences over a specified number of epochs, denoted as $k$. The critic network’s parameters are updated by minimizing the critic loss function, denoted as $L_c$, which aims to minimize the loss over the sampled mini-batch of experiences.

\RestyleAlgo{ruled}
\SetKwComment{Comment}{/* }{ */}
\begin{algorithm}
\caption{Proximal Policy Optimization Algorithm}\label{alg:ppo}
Initialize the parameter of the actor-critic network \\
\While{termination criteria is not satisfied}{
    Step 1: Generate N experiences
    $ \left\{\mathrm{s}_{t_1}^{}, \mathrm{a}_{t_1}^{}, \mathrm{r}_{t_1}^{}\right\}, \left\{\mathrm{s}_{t_2}^{}, \mathrm{a}_{t_2}^{}, \mathrm{r}_{t_2}^{}\right\}... \left\{\mathrm{s}_{t_n}^{}, \mathrm{a}_{t_n}^{}, \mathrm{r}_{t_n}^{}\right\}$\;
    Step 2: Calculate action-value function and advantage function at each time step t,
    $$ \mathrm{Q}_{}^{\pi}(s,a) =  \sum_{t}^{}\mathrm{E}_{\pi\theta}^{}[R(s_t,a_t)|s,a]$$
     $$\mathrm{V}_{}^{\pi}(s) =  \sum_{t}^{}\mathrm{E}_{\pi\theta}^{}[R(s_t,a_t)|s]$$
     $$\mathrm{A}_{}^{\pi}(s,a) =  \mathrm{Q}_{}^{\pi}(s,a)-\mathrm{V}_{}^{\pi}(s)$$
  \While{$k \not = epoch$}{
    1. The critic network’s parameters are updated by:
    $$\mathrm{L}_{c}^{}({\theta}_v) = \frac{1}{M}\sum_{t=1}^{M}(\mathrm{Q}_{}^{\pi}(s,a)-V(s|\theta_v))^2$$
    2. The actor network’s parameters are updated by:
    $$\mathrm{L}_{a}^{}(\theta_v)= -\frac{1}{M}[min(r_t(\theta)A_t,clip(r_t(\theta),1-\epsilon,1+\epsilon)A_t)]$$
  }
}
\end{algorithm}
On the other hand, the actor network’s parameters are updated by repeating a series of steps until a 
terminating criterion is met. These steps involve maximizing a surrogate objective function that balances the exploration and exploitation trade-off. The objective is to find the policy that maximizes 
the expected reward. The training process continues iteratively, with the critic and actor networks 
being updated based on their respective loss functions and objectives. This iterative process allows the PPO algorithm to progressively refine the actor-critic networks and improve the overall decision-making capabilities. The training process persists until a predetermined termination criterion, such as 
attaining a maximum number of iterations or accomplishing a specified level of performance, is 
fulfilled. At this point, the PPO algorithm concludes its training process.
\subsection{Reward Calculation Process for the RL Agent}
The subsequent crucial step is the computation of rewards, which accurately assesses the accuracy of 
the action generated by the RL agent. During the training of the RL agent, a fixed number of sample steps are used unless the termination criterion is met. If the output voltage of the boost converter
exceeds the upper limit ($V_{up}$) or is greater than $V_{ref}$ and then crosses the lower limit ($V_{low}$), the training for that episode is terminated. This logic can be considered as a limiting mechanism that helps to reach and stay at the desired voltage level. At each sample time step ($t$) during training, the RL agent takes the state, ($s_{t-1}$) and the reward value, ($r_{t-1}$) as inputs to generate the next action, as shown in Fig. \ref{fig:2}.
The reward function, specified in Algorithm \ref{alg:reward}, determines the reward value based on the current state 
and desired outcomes, guiding the agent’s learning process.
\begin{algorithm}[ht]
\caption{Reward calculation}
\label{alg:reward}
\DontPrintSemicolon
 \begin{algorithmic}
    \\
        \State $V_{out} \gets  \text{Output Voltage}$
        \State $V_{ref} \gets \text{Desired Output (Reference)}$
        \State $e_{th} \gets \text{Threshold Error}$
        \State $V_{up} \gets \text{Upper limit of voltage - } V_{up}=(1+ e_{th} ).V_{ref}$
         \State $V_{low} \gets \text{Lower limit of voltage - } V_{low}=(1-e_{th} ).V_{ref}$
     \\
 \end{algorithmic}

  \If{$V_{out} \geq V_{ref}$}
    {
        flag = 1;
    }
  \If{$(V_{out} \geq V_{up}) \parallel (V_{out} \leq V_{low} \&\&  \text{flag}==1$)}
    {
        $r(t) = -1;$\\
   \Else
        {
        $r(t) = \frac{1}{abs(e(t))};$
        }
    }

\end{algorithm}

\section{Simulation result and analysis}
In this section, the simulation results using the proposed method are described. Table \ref{tab_1} shows the design specification of the boost converter used for this work. The simulations are operated for two different conditions, each of the conditions consists of three different scenarios, and results are compared with different existing control methods with different scenarios. The justification behind integrating the reinforcement learning technique to regulate the dc-to-dc boost converter is verified by an in-depth analysis of each situation. Table \ref{tab_2} provides a summary of the simulation conditions and scenarios.

\begin{table}[h]
\caption{Boost converter design specification}
\label{tab_1}
\centering
\normalsize 
\def\arraystretch{1.25}
\begin{tabular}{lc}
\hline
\textbf{Design Parameter} & \textbf{Values}  \\
\hline
Input Voltage, $V_{in}$  & 24V    \\
Desired Output Voltage, $V_{ref}$ & 48/54/60V \\
Load Resistance, $R$ & 50$\Omega$\\
Inductor, $L$ & 10$\mu$H  \\
Capacitor, $C$ & 400mF \\
\hline
\end{tabular}
\vspace{8pt}
\end{table}


\begin{table}[]
\centering
\caption{Simulation conditions and scenarios.}
\label{tab_2}
\begin{tabular}{cc}
\hline
Conditions                                      & Scenarios \\ \hline
\multirow{3}{*}{I. Fixed Input (24 V)}          & Ref-48 V  \\  
                                                & Ref-54 V  \\ 
                                                & Ref-60 V  \\ \hline
\multirow{3}{*}{II. Variable Input (24 V–26 V)} & Ref-48 V  \\ 
                                                & Ref-54 V  \\ 
                                                & Ref-60 V  \\ \hline
\end{tabular}
\end{table}

\begin{figure}[h]
    \centering
    \includegraphics[width=0.8\linewidth]{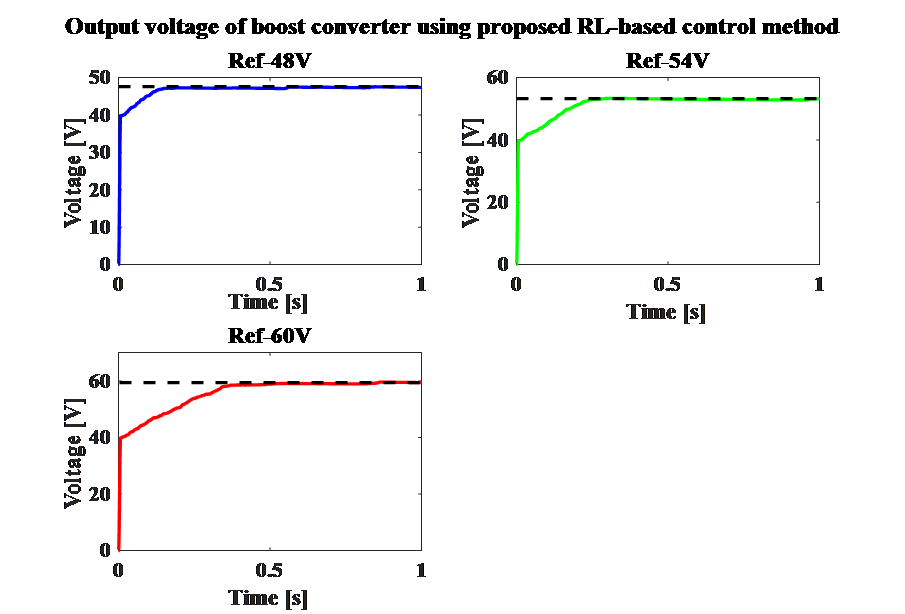}
    \caption{Output voltage of boost converter using proposed control method (Condition I)}
    \label{fig_4}
\end{figure}

\begin{table}[]
\centering
\caption{Step response characteristics of RL control method (Condition I)}
\label{tab_3}
\begin{tabular}{cccc}
\hline
Step Response Characteristics & Ref 48 V & Ref 54 V & Ref 60 V \\ \hline 
Rise time (sec)               & 0.056    & 0.135    & 0.239    \\ 
Settling time (sec)           & 0.123    & 0.218    & 0.360    \\ 
Overshoot (\%)                & 0.364    & 0.367    & 0.275    \\ 
Undershoot (\%)               & $7.3\times 10^-7$  & $6.5 \times 10^-7$ & $5.8\times 10^-7$ \\ \hline
\end{tabular}
\end{table}

\subsection{Conditions and simulation scenarios}
The experimentation is conducted under two distinct conditions to evaluate the performance of the proposed method. In the first condition, a constant input voltage of 24 V is maintained. This specific setting allows for the observation and assessment of the control capability inherent in the traditional application of a boost converter. On the other hand, the second condition involves a dynamic scenario where the input voltage fluctuates between 24 V and 26 V. The primary objective here is to scrutinize the controller’s efficacy in managing the variable input behavior. Notably, to induce the input variation, a step change is introduced precisely at 0.5 s of the simulation. This comprehensive experimental design enables a thorough examination of the proposed method’s adaptability and control performance under both stable and dynamic operating conditions. For each condition, three scenarios are being taken into account. Each of those scenarios corresponds to a distinct reference voltage (48 V, 54 V, and 60 V). The objective of this experiment is to evaluate whether the suggested controller can achieve the desired voltage level while exhibiting precise step response characteristics.
\begin{figure}[h]
    \centering
    \includegraphics[width=0.8\linewidth]{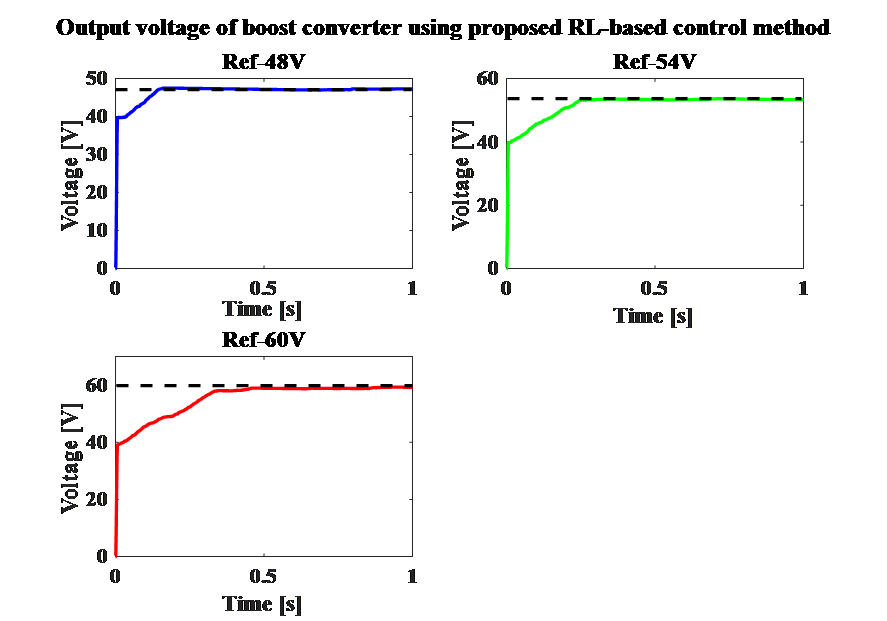}
    \caption{ Output voltage of boost converter using proposed control method (Condition II)}
    \label{fig_5}
\end{figure}

\begin{table}[]
\centering
\caption{Step response characteristics of RL control method (Condition II)}
\label{tab_4}
\begin{tabular}{cccc}
\hline
Step Response Characteristics & Ref 48 V & Ref 54 V & Ref 60 V \\ \hline 
Rise time (sec)               & 0.093    &  0.155    &  0.259    \\ 
Settling time (sec)           & 0.163    &  0.274    & 0.395    \\ 
Overshoot (\%)                &  0.347    & 0.056    & 0.161    \\ 
Undershoot (\%)               & 0.000  & 0.000 & 0.000 \\ \hline
\end{tabular}
\end{table}
\subsection{Result analysis}
Condition I (Fixed Input Voltage): In the case of fixed input voltage, the performance of the proposed control method is shown in Fig. \ref{fig_4}, and the step response characteristics are described in Table \ref{tab_3}. It is quite evident that the proposed controller is capable of regulating the constant voltage at the output end with a good step response characteristic. \\
Condition II (Varying Input Voltage): In the event of varying input voltage, the performance of the proposed control method can be observed in Fig. \ref{fig_5}, and the step response characteristics are detailed in Table \ref{tab_4}. The suggested controller is also capable of maintaining a constant voltage at the output, exhibiting a favorable step response characteristic, even when the input voltage varies.

\section{Validation of proposed method and performance comparison}
\subsection{Existing methods}
\subsubsection{PI Control}
\begin{figure}[h]
    \centering
    \includegraphics[width=0.8\linewidth]{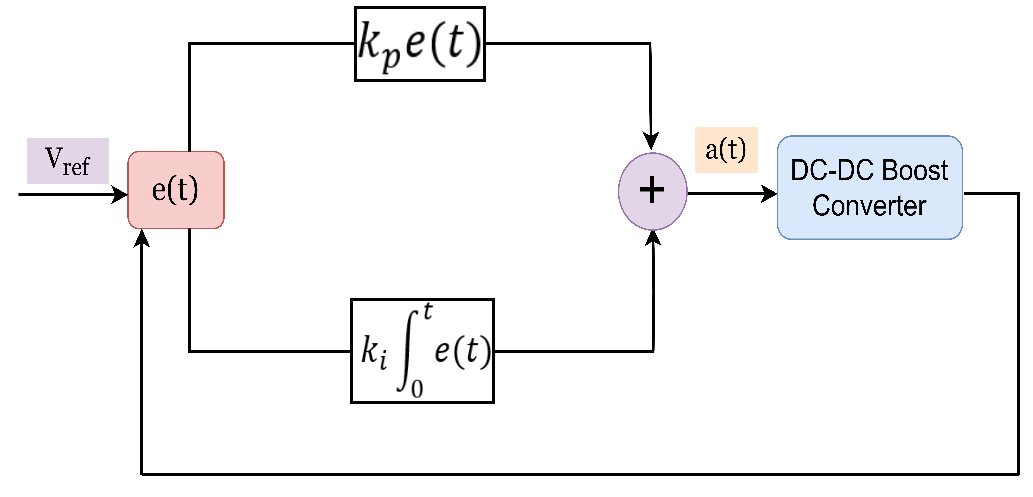}
    \caption{Boost converter with PI control}
    \label{fig_6}
\end{figure}

The PI controller is a feedback control loop that calculates an error signal by measuring the difference between the system’s output 
and a desired reference value. This algorithm constantly examines the output voltage and modifies the duty cycle of the converter to uphold the desired output voltage level. To put it simply, it closely monitors the output voltage and adjusts it as needed to maintain the 
correct level. The PI controller utilizes a feedback control loop mechanism to minimize the influence of disturbances in a system, steer the system 
towards a desired state, and define explicit linkages between system variables. The input of the system is the error at a certain time, $e(t)$, which represents the difference between the measured and reference values as shown in Fig. \ref{fig_6}. The PI controller produces an 
output called ‘actuation’, denoted as $a(t)$, which determines the action to be implemented for the given plant or system. The actuation, $a(t)$, is determined by adding two components: the product of the proportional gain ($k_p$) and the magnitude of the error, and the product of the integral gain ($k_i$) and the integral of the error over time. The function, $a(t)$, can be written as Eq. \ref{eq:pi}, 

\begin{equation}
\label{eq:pi}
a(t)=k_p*e(t)+k_i\int_{0}^{t}e(t)dt
\end{equation}
The PI controller combines proportional and integral terms to generate a control signal, $a(t)$, at time $t$. The proportional term ($k_p$) 
responds to the current error, reducing steady-state error. However, relying solely on proportional gain can lead to oscillations. To 
mitigate this, the integral term ($k_i$) considers accumulated past errors, aiding in the gradual reduction of steady-state errors over time. 
To perform properly, the PI controller parameters have to be appropriately selected. For this selection process in this work, two highly 
effective algorithms are used namely, the particle swarm algorithm (PSO) \cite{juneja2016particle, solihin2011tuning} and the genetic algorithm (GA) \cite{meng2007fast, katoch2021review}. Leveraging PSO, the goal is to optimize $k_p$ and $k_i$ for efficient and accurate boost converter control. Initiating with the definition of the objective function as mean absolute error (MAE), it serves as the metric for evaluating solution quality. With two variables ($k_p$ and $k_i$), search space is constrained by lower and upper bounds. Implementing PSO in MATLAB, the ‘particleswarm’ function is utilized, 
specifying parameters such as swarm size, maximum iterations, display options, and plot function. Executed by invoking ‘particleswarm’ with the objective function, variables, bounds, and designated options, the algorithm yields optimized $k_p$ and $k_i$ values assigned to specific parameters.

The genetic algorithm (GA) is also used for validation to optimize the values of two control parameters, $k_p$, and $k_i$, which are critical 
for regulating a DC-DC boost converter. Configured with specific parameters and constraints, the GA uses the objective function MAE to guide optimization. The optimization process is initiated by invoking the ‘ga’ function in MATLAB, specifying the objective function, 
number of variables, and variable bounds. The resulting optimized solution provides values for $k_p$ and $k_i$.

\subsubsection{ANN Control}
\begin{figure*}[h]
    \centering
    \includegraphics[width=0.8\linewidth]{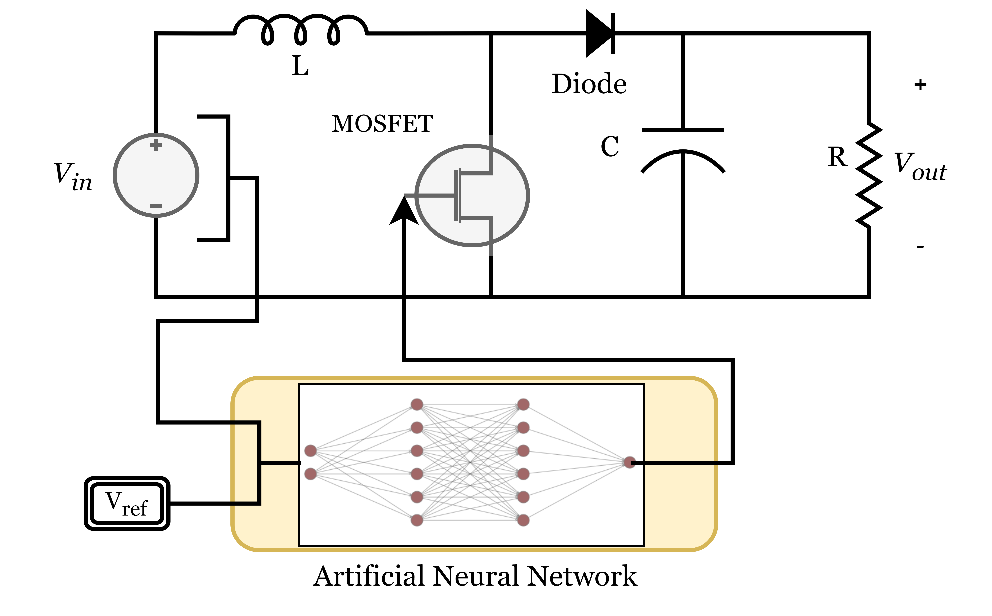}
    \caption{Boost converter with ANN control}
    \label{fig_7}
\end{figure*}

A DC-DC boost converter with artificial neural network (ANN) control is a power converter that enhances the voltage of a DC input source by utilizing an advanced control mechanism based on neural network technology. The ANN control mechanism mimics the behavior of a human neural network, enabling the converter to adapt and enhance its performance in real time. The ANN control algorithm continuously checks both the converter’s input, $V_{in}$, and the target output, $V_{ref}$ to ensure optimal performance as shown in  Fig. \ref{fig_7}, altering its control signals accordingly.

A substantial amount of data (100000 data samples) is generated using the MATLAB editor during the development of an artificial neural network (ANN) model, generating the dataset. The dataset is made up of data samples with input and output voltages that serve as input features for the ANN model. 80\% of the data is used for training, and the remaining 20\% is used for testing. Following that, the neural network is built and trained using MATLAB’s graphical user interfaces (GUIs) intended exclusively for neural network (NN) applications. After the dataset has been built within the MATLAB workspace, these GUIs make it easier to create and train the neural network. To improve the performance of the neural network (NN) model, the training parameters were fine-tuned. Table \ref{tab_5} shows the specifications of the NN model. After training, the NN model is exported to the MATLAB workspace and converted into a Simulink block. This Simulink block manages the boost 
converter’s duty cycle, giving it control over its operation.

\begin{table}[]
\centering
\caption{Specifications of neural network.}
\label{tab_5}
\begin{tabular}{cc}
\hline
Parameters                 & Training Configuration       \\ \hline
Type of Network            & Feed-Forward Backpropagation \\ 
Training Function          & TRAINLM                      \\ 
Adaption Learning Function & LEARNGDM                     \\ 
Objective Function         & MSE                          \\ 
Number of Layers           & 3                            \\ 
Transfer Function          & TANSIG                       \\ \hline
\end{tabular}
\end{table}

\subsection{Performance analysis}
In this subsection, a comparative analysis between the proposed method and existing methods is depicted. The comparison is 
shown for both conditions with three individual scenarios.

\begin{figure*}[h]
    \centering
    \includegraphics[width=0.8\linewidth]{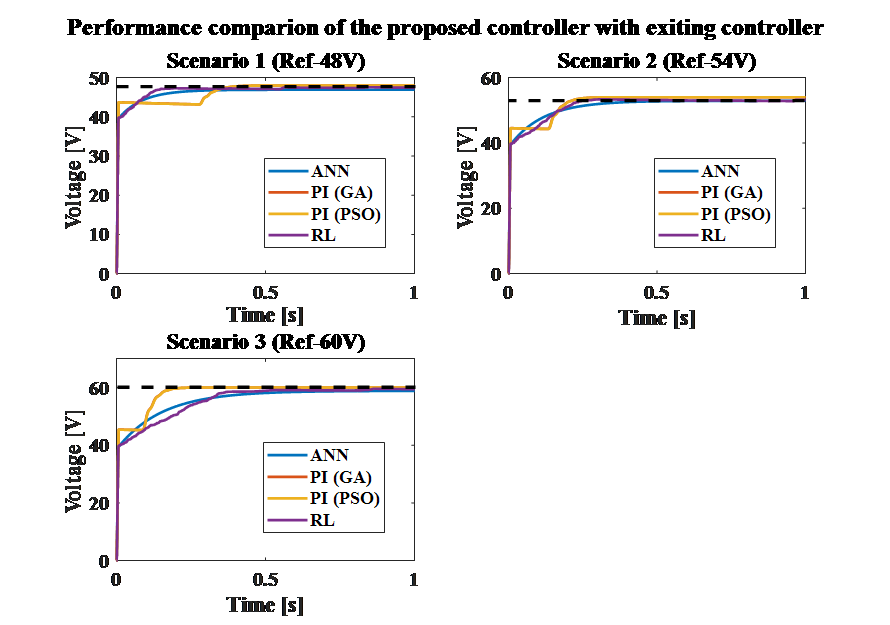}
    \caption{Step response characteristics comparison (Condition I)}
    \label{fig_8}
\end{figure*}

\noindent\textbf{Condition I. }When the input voltage remains constant, the optimized PI controller consistently achieves the shortest settling time compared to other control methods. Interestingly, for the PI controller, the settling time decreases as the desired voltage increases as shown in Fig. \ref{fig_8}. The quantitative information on the settling time for different control methods under fixed input voltage conditions is presented in Table \ref{tab_6}. The optimized PI controller consistently exhibited the shortest settling time across various scenarios, with the settling time decreasing as the desired voltage increased. This trend indicates the controller’s ability to achieve faster stabilization as the system becomes more aggressive in its response. This behavior can be attributed to the inherent dynamics and characteristics of the system being controlled. This could be due to the system becoming more responsive and active as the desired voltage increases.
 
However, it’s important to note that this behavior may not be the same in all cases and can depend on the system’s characteristics and control parameters chosen. Different trends are observed when using RL and ANN control methods as depicted in Fig. \ref{fig_8}. Among the RL and ANN control methods, the boost converter with RL control performed better in terms of settling time which is quite evident in Table \ref{tab_6}. This is because RL has unique learning and decision-making abilities that allow it to adapt and optimize the control policy according to the specific needs and dynamics of the boost converter system. 

\begin{figure*}[h]
    \centering
    \includegraphics[width=0.8\linewidth]{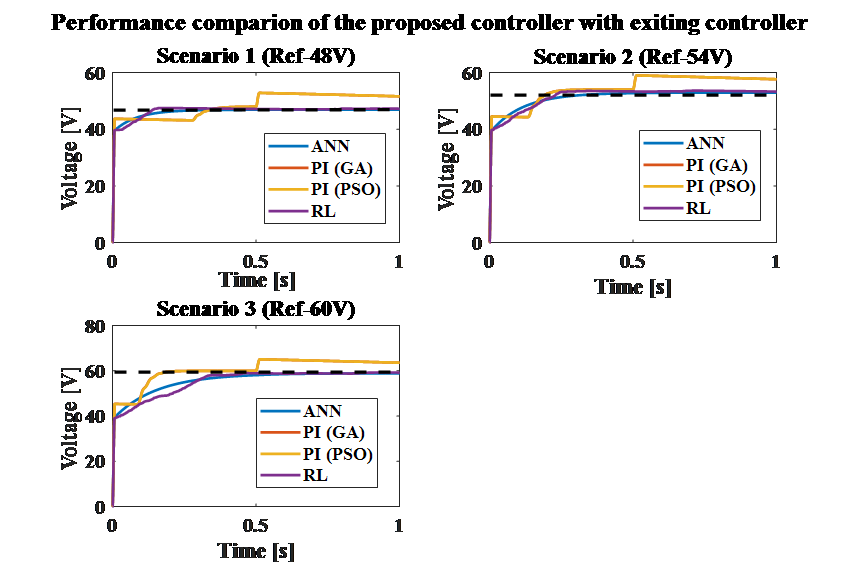}
    \caption{ Step response characteristics comparison (Condition II)}
    \label{fig_9}
\end{figure*}

When the input voltage is varied, the step response characteristics of the system are observed to remain relatively consistent for both ANN control and RL control methods depicted in Fig. \ref{fig_9}. This implies that these methods are capable of adapting to the varying input voltage and maintaining stable step response characteristics. However, in the case of PI control, undesired voltage wave shapes are evident in Fig. \ref{fig_9}. The step response characteristics are 
significantly degraded, indicating that the PI control method struggled to handle the input voltage variation effectively. Quantitatively comparing the performance of these control methods, RL control emerged as the superior method as depicted in Table \ref{tab_7}. It exhibited the ability to seamlessly handle the input voltage variation and maintain step response characteristics similar to those observed under fixed input voltage conditions. This quality positions RL control as the best control method among the three in terms of adapting to varying input voltage and preserving stable step response characteristics. The RL control method’s effectiveness can be attributed to its capacity to learn and optimize the control policy based on the specific dynamics and requirements of the boost converter system. This adaptability allows RL-based control to consistently deliver reliable and robust performance, even in the presence of changing input voltage conditions.


\section{Discussion}
\subsection{Duty of the proposed method}
The primary duty of the proposed PPO-based reinforcement learning (RL) method in the experimental setup is to improve the 
control performance of a DC-DC boost converter. The key duties and outcomes observed in the experiments are as follows:
\begin{itemize}
    \item The PPO-based RL method provided faster settling times and improved stability metrics, indicating a more responsive and stable 
    control system.
    \item The proposed method was tested under varying input voltage condition, where it demonstrated robust performance without significant degradation, thus validating its applicability in dynamic environments.
    \item Throughout the training and testing phases, safety constraints were incorporated to ensure that the control actions remained within safe operating limits. The method successfully avoided unsafe states, thereby ensuring reliable and safe operation.
\end{itemize}
\subsubsection{Requirements of the computational speed of controller}
The proposed PPO-based reinforcement learning (RL) method for DC-DC boost converter control does have slightly higher computational demands compared to traditional control methods. This is primarily due to the complexity of the PPO algorithm and the 
need for real-time data processing and decision-making. However, the increasing availability and affordability of powerful computational hardware, such as GPUs and specialized processors, mitigate this concern. Additionally, further optimization of the PPO algorithm can reduce its computational load, making it more suitable for real-time applications. In practical implementations, careful selection of hardware and algorithmic optimizations can ensure that the computational requirements are met without compromising 
performance.
\subsubsection{Safety of the training process}
Ensuring the safety of the training process for the PPO-based RL method is important, especially when dealing with power elec
tronics like DC-DC converters. The following strategies can be employed to guarantee safety:
\begin{itemize}
    \item Conducting the majority of the training in a high-fidelity simulated environment (such as MATLAB Simulink) can prevent any risk 
    to physical components during the initial learning phase. This allows the algorithm to learn and adapt without causing harm to 
    actual hardware.
    \item Incorporating safety constraints within the RL framework to ensure the actions taken by the algorithm do not lead to unsafe 
    operating conditions. These constraints can be designed to limit voltage, current, and other critical parameters within safe ranges.
    \item The trained model can be deployed gradually into the physical system. The system can be started with limited operational scenarios and progressively increase in complexity as it demonstrates a stable and safe response.
\end{itemize}
By following these strategies, the safety of the training process can be effectively managed, ensuring reliable and secure deployment of the PPO-based RL method in practical applications.

\section{Limitations}
While the proposed PPO-based reinforcement learning (RL) approach for DC-DC boost converter control shows significant promise, 
a few limitations should be noted.
\begin{itemize}
    \item The computational complexity of the proposed RL-based control method is slightly higher than that of traditional control methods. 
    However, with the ongoing advancements in computational hardware, this limitation is becoming less significant. Future research 
    can focus on optimizing the algorithm to reduce its computational requirements.
    \item The performance of the proposed controller under varying operating conditions and disturbances needs further exploration. 
Although the current study shows positive results in controlled simulation environments, real-world conditions may present additional challenges. Future work can include extensive testing under diverse and unpredictable scenarios to enhance the algorithm’s robustness.
\item While this research is focused on DC-DC boost converters, the adaptability of the proposed PPO-based RL approach to other types of 
converters or applications has not been extensively tested. Future studies can extend this research to a broader range of applications 
to confirm the generalizability of the method.
\end{itemize}
By addressing these limitations in future research, the effectiveness and applicability of the proposed PPO-based RL approach can 
be further improved.

\begin{sidewaystable}[]
\small
\centering
\caption{Step response characteristics (Condition 1).}
\label{tab_6}
\begin{tabular}{>{\centering\arraybackslash}p{2cm} >{\centering\arraybackslash}p{3cm} >{\centering\arraybackslash}p{3cm} >{\centering\arraybackslash}p{3cm} >{\centering\arraybackslash}p{3cm}}
\toprule
 & ANN & PI (GA) & PI (PSO) & RL (Proposed) \\ \cmidrule(lr){2-2} \cmidrule(lr){3-3} \cmidrule(lr){4-4} \cmidrule(lr){5-5}
 & 48 V \quad 54 V \quad 60 V & 48 V \quad 54 V \quad 60 V & 48 V \quad 54 V \quad 60 V & 48 V \quad 54 V \quad 60 V \\ \midrule
Rise Time (s) & 0.04 \quad 0.11 \quad 0.14 & 0.005 \quad 0.160 \quad 0.120 & 0.005 \quad 0.154 \quad 0.120 & 0.056 \quad 0.135 \quad 0.239 \\ 
Settling Time (s) & 0.170 \quad 0.290 \quad 0.420 & 0.350 \quad 0.213 \quad 0.163 & 0.343 \quad 0.212 \quad 0.162 & 0.123 \quad 0.218 \quad 0.360 \\ 
Overshoot (\%) & 0 \quad 0 \quad 0 & $3.6 \times 10^{-5}$ \quad $5.6 \times 10^{-5}$ \quad $4.83 \times 10^{-5}$ & $5.13 \times 10^{-5}$ \quad $5.39 \times 10^{-5}$ \quad $4.59 \times 10^{-5}$ & 0.364 \quad 0.367 \quad 0.275 \\ 
Undershoot (\%) & $1.27 \times 10^{-5}$ \quad $1.25 \times 10^{-5}$ \quad $1.22 \quad 10^{-5}$ & $2.5 \times 10^{-5}$ \quad $2.5 \times 10^{-5}$ \quad $2.5 \times 10^{-5}$ & $2.52 \times 10^{-5}$ \quad $2.52 \times 10^{-5}$ \quad $2.52 \times 10^{-7}$ & $7.3 \times 10^{-5}$ \quad $6.51 \times 10^{-5}$ \quad $5.83 \times 10^{-5}$ \\ 
\bottomrule
\end{tabular}
\end{sidewaystable}

\begin{sidewaystable}[]
\centering
\caption{Step response characteristics (Condition II).}
\label{tab_7}
\begin{tabular}{cccccc}
\toprule
 & ANN & PI (GA) & PI (PSO) & RL (Proposed) \\ \cmidrule(lr){2-2} \cmidrule(lr){3-3} \cmidrule(lr){4-4} \cmidrule(lr){5-5}
 & 48 V \quad 54 V \quad 60 V & 48 V \quad 54 V \quad 60 V & 48 V \quad 54 V \quad 60 V & 48 V \quad 54 V \quad 60 V \\ \midrule
Rise Time (s) & 0.041 \quad 0.110 \quad 0.186 & 0.330 \quad 0.190 \quad 0.140 & 0.327 \quad 0.190 \quad 0.142 & 0.056 \quad 0.135 \quad 0.239 \\ 
Settling Time (s) & 0.179 \quad 0.296 \quad 0.435 & 0.600 \quad 0.590 \quad 0.590 & 0.590 \quad 0.590 \quad 0.590 & 0.123 \quad 0.218 \quad 0.360 \\ 
Overshoot (\%) & 0 \quad 0 \quad 0 & 2.350 \quad 2.250 \quad 2.350 & 2.350 \quad 2.350 \quad 2.350 & 0.364 \quad 0.367 \quad 0.275 \\ 
Undershoot (\%) & 0 \quad 0 \quad 0 & 0 \quad 0 \quad 0 & 0 \quad 0 \quad 0 & 0 \quad 0 \quad 0 \\ 
\bottomrule
\end{tabular}%
\end{sidewaystable}


\section{Summary and Outlook}
In this research, we introduced an effective control strategy for DC-DC boost converters based on proximal policy optimization 
(PPO), a reinforcement learning (RL) algorithm. Our approach addresses the limitations inherent in traditional control methods, of
fering improved performance in controlling boost converters. To validate the effectiveness of our proposed method, we conducted a 
comparative analysis against conventional control techniques. The results showed that the RL-based control method performed more 
consistently than other methods when the inputs were fixed or changed, particularly in terms of step response characteristics. In 
contrast, the PI control methods, optimized using particle swarm optimization (PSO) and genetic algorithm (GA), exhibited comparable performance, while the ANN-based control system also maintained satisfactory results, highlighting the potential of artificial 
intelligence in converter control. Overall, our findings suggest that the PPO-based RL approach is the most effective among the evaluated methods, consistently delivering superior step response characteristics across varying input voltages. Despite some limitations, our approach shows significant promise. Future research should focus on addressing these limitations to further enhance the effectiveness and applicability of this RL-based control method in DC-DC boost converters.

\bibliography{sn-bibliography}

\end{document}